\documentclass[sigconf]{acmart}

\AtBeginDocument{%
  \providecommand\BibTeX{{%
    \normalfont B\kern-0.5em{\scshape i\kern-0.25em b}\kern-0.8em\TeX}}}

\copyrightyear{2020}
\acmYear{2020}
\setcopyright{acmcopyright}\acmConference[CIKM '20]{Proceedings of the 29th ACM International Conference on Information and Knowledge Management}{October 19--23, 2020}{Virtual Event, Ireland}
\acmBooktitle{Proceedings of the 29th ACM International Conference on Information and Knowledge Management (CIKM '20), October 19--23, 2020, Virtual Event, Ireland}
\acmPrice{15.00}
\acmDOI{10.1145/3340531.3412115}
\acmISBN{978-1-4503-6859-9/20/10}

\usepackage{graphicx}
\usepackage{subfigure}
\usepackage{bbm}
\usepackage{booktabs}
\usepackage{multirow}

\begin{document}
\fancyhead{} 
\title{DREAM: A Dynamic Relational-Aware Model for Social Recommendation}

\author{Liqiang Song, Ye Bi, Mengqiu Yao, Zhenyu Wu, Jianming Wang, Jing Xiao}
\email{magicyebi@163.com, {songliqiang537, yaomengqiu621, wuzhenyu447, wangjianming888, xiaojing661}@pingan.com.cn}
\affiliation{%
\institution{Ping An Technology Shenzhen Co., Ltd}
}



\begin{abstract}
Social connections play a vital role in improving the performance of recommendation systems (RS). However, incorporating social information into RS is challenging. Most existing models usually consider social influences in a given session, ignoring that both users’ preferences and their friends’ influences are evolving. Moreover, in real world, social relations are sparse. Modeling dynamic influences and alleviating data sparsity is of great importance.

In this paper, we propose a unified framework named Dynamic RElation-Aware Model (DREAM) for social recommendation, which tries to model both users’ dynamic interests and their friends’ temporal influences. Specifically, we design temporal information encoding modules, because of which user representations are updated in each session. The updated user representations are transferred to relational-GAT modules, subsequently influence the operations on social networks. In each session, to solve social relation sparsity, we utilize glove-based method to complete social network with virtual friends. Then we employ relational-GAT module over completed social networks to update users’ representations. In the extensive experiments on the public datasets, DREAM significantly outperforms the state-of-the-art solutions.

\end{abstract}





\maketitle



\section{Introduction}

Recommendation systems (RS), attempting to help users overcome information overload, have become popular. RS make use of users' historical activities and/or social relationships to generate features of items and identify users' latent preferences. Among RS strategies, collaborate filtering (CF)-based methods have received significant success \cite{DBLP:conf/uai/RendleFGS09}. However, CF-based methods usually suffer from sparsity of user-item interactions and cold start problem. 

To address these limitations, modern RS also generate other useful data. By combining rating matrices with additional data, better recommendations are respected. Nowadays, the increasing popularity of social media greatly enriches people's social activities, which harnesses social relations to boost the performance of RS \cite{DBLP:conf/www/Fan0LHZTY19}. Based on the intuition that people in same social group are likely to have similar preferences, and that users will gather information from their social friends, lots of work has incorporated social information into RS, e.g. SBPR \cite{DBLP:conf/cikm/ZhaoMK14}, GraphRec \cite{DBLP:conf/www/Fan0LHZTY19}, etc..However, incorporating social information into RS is challenging.

In general, methods utilizing heterogeneous data tend to perform better than those using a single data source \cite{DBLP:conf/cikm/KangPKCC19}. 
DGRec \cite{DBLP:conf/wsdm/Song0WCZT19} models dynamic user behaviors with RNN, and social influence with GAT, and is proven to outperform the state-of-the-art solutions. However, the work only considers social influences in a given session, ignoring the effects of temporal dependency among different sessions. Besides, relying only on social friends is far from satisfactory, since in real world, the social relations are very sparse.

Based on the observations, we propose a novel framework called A Dynamic RElational-Aware Model (DREAM), which tries to model both users’ dynamic individual interests and their friends’ temporal influences. The model contains two main modules, temporal information encoding (TIE) modules encode the outputs form historical sessions by recursively combining the representation of all previous outputs with the current output. In each session, we first complete social network, and then employ relational-aware graph attention (relational-GAT) network modules to integrate influences from users' real and virtual friends. To be specific, we first design a method to increase the number of friends, which refer to the neighbors whose behaviors are similar to the users, and we call this kind of neighbors as virtual friends. Within each session, we then utilize relational-GAT \cite{DBLP:conf/cikm/XuLHLX019} to capture the influences of friends. 
As we know, user preference for products drifts over time. To model users' dynamic preferences and the dynamic influences from their friends, we design the temporal information encoding modules. In each TIE module, we combine the relational-GAT encoded features as well as the features from last TIE module. 
In summary, our contributions in this paper are as follows:
\begin{itemize}
\item 
We propose a novel RS approach, which aims to model users’ dynamic interests and dynamic influences from their friends.  The model encodes the outputs form historical sessions by recursively combining the features encoded by relational-GAT modules and that from last TIE module.
\item We design a GloVe-based method to increase the number of friends, and use relational-GAT to aggregate user representations from both completed social network in each session. 
\item We conduct experiments on real-world recommendation scenarios, and the results prove the efficacy of DREAM over several state-of-the-art baselines.
\end{itemize}


\section{Problem Formulation}

Let $\mathcal{U}=\{u_{1},u_{2},\ldots u_{n}\}$ and $\mathcal{I}=\{i_1, i_2, \ldots i_{m}\}$ denote the sets of users and items, where $n$ and $m$ are the number of users and items, respectively.
The user-item interaction matrix $\boldsymbol{Y}\in\mathbb{R}^{n\times m}$
is defined as: $y_{u,i}=1$, if there is an interaction between $u$ and $i$, and $0$ otherwise.
Each $u$ is associated with a set of sessions, $\mathcal{I}_{T}^{u}=\{\mathcal{S}_{1}^{u}, \mathcal{S}_{2}^{u}, \ldots, \mathcal{S}_{T}^{u}\}$, where $\mathcal{S}_{t}^{u}$ is the $t$-th session of user $u$, each session $\mathcal{S}_{t}^{u}$ consists of a user behaviors sequence $\{i_{t,1}^{u}, i_{t,2}^{u}, \ldots, i_{t, N_{u,t}}^{u}\}$, where $i^{u}_{t,p}$ is the $p$-th item interacted by user $u$ in $t$-th session. Social network can be described by $\boldsymbol{S}^{R}\in\mathbb{R}^{n\times n}$, where $s^{R}_{p,q}=1$ if there is a relation between $u_p$ and $u_q$, and $0$ otherwise. 
We call embedding vectors $\boldsymbol{u}$ and $\boldsymbol{i}$ latent feature vectors. Given $\boldsymbol{Y}$ and $\boldsymbol{S}^{R}$, we aim to predict whether $u$ has potential interests in target item $v$.


\section{DREAM Framework}

\label{www19_construct}

\begin{figure}[!h]
\setlength{\abovecaptionskip}{0cm}
\setlength{\belowcaptionskip}{0cm}
  \centering
  \includegraphics[scale=0.38]{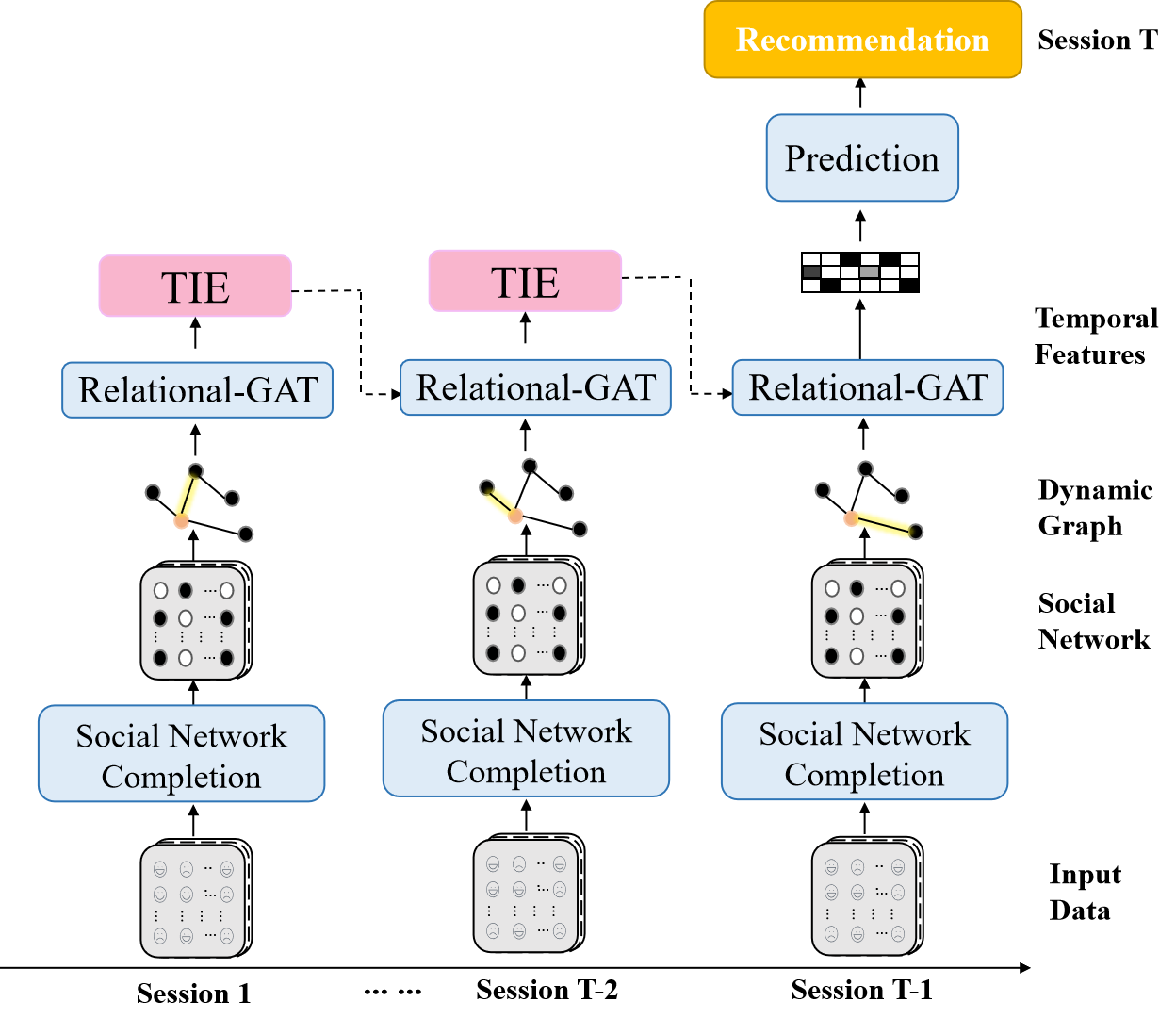}
  \caption{DREAM Framework. In each session, relational is employed to integrate influences from completed social network. User representations are updated in TIE module and transferred to next relational-GAT module.}
  \label{www2020_framework}
\end{figure}

The framework is illustrated in Figure \ref{www2020_framework}. The model consists of social network completion, relational-aware graph attention network (relational-GAT) modules, temporal information encoding (TIE) modules and recommendation. 
Given a user $u$, we first select her $T$ historical sessions. In $t$-th session, we first complete social network, and get $\mathcal{G}_{t}^{C}$.
The inputs of the model are graph sequence $\{\mathcal{G}_{1}^{C}, \mathcal{G}_{2}^{C},\ldots, \mathcal{G}_{T}^{C}\}$. In each session, relational is employed to integrate influences from $\mathcal{G}_{t}^{C}$. User representations are updated in TIE module and transferred to next relational module.

\subsection{Social Network Completion}

{\bf Virtual Friends Definition and Selection}. 
We define \emph{virtual friend} as users having similar consumption habits, and the connection is stronger if they are more similar. Under the definition, we design a GloVe-based method. We first utilize GloVe mechanism \cite{DBLP:conf/emnlp/PenningtonSM14} to learn user representations, and then calculated the similarity among all users, finally we choose top-k users whose similarity is higher. The input is user-user co-occurrence counts matrix $\boldsymbol{X}$, whose entries $X_{p,q}$ denote the number of times user $p$ and user $q$ consume the same items. And outputs are user embeddings $\boldsymbol{g}_{u}$.
We define the connection strength between user $u_{p}$ and $u_{q}$ as:
\begin{eqnarray*}
\label{WWW19_vscore}
s_{p,q}^{V} = \text{softmax}(\langle\boldsymbol{g}_{u_p},\boldsymbol{g}_{u_q}\rangle) = \frac{\exp(\langle\boldsymbol{g}_{u_p},\boldsymbol{g}_{u_q}\rangle)}{\sum_{u_{l},u_{s}\in\mathcal{U}}\exp(\langle\boldsymbol{g}_{u_l},\boldsymbol{g}_{u_s}\rangle)},
\end{eqnarray*}
and $u_{p}$'s virtual friends set is defined as $\mathcal{N}^{V}(u_{p})$.

\noindent{\bf Node Representation}. 
Friends' influences always lag, since friends may consumed products first and then influence the user. So in $t$-th relational-GAT module, friends node representations are calculated in ($t$-1)-th session. Here, we use users' short-term interests, which are gotten by employing GRU \cite{DBLP:journals/corr/HidasiKBT15} over user $u_{j}$'s interaction sequence in ($t$-1)-th session $\mathcal{S}_{t-1}^{j}=\{i_{t-1,1}^{j}, i_{t-1,2}^{j}, \ldots, i_{t-1, N_{k,t}}^{j}\}$, i.e. $\boldsymbol{s}_{j} = \text{GRU}(\mathcal{S}_{t-1}^{j})$.


\subsection{Relational-GAT Module}

\label{www20_graph_atten}
Note there are two different relations in our completed social network, to capture this difference, we employ relational-GAT over the completed social network in each session. In $t$-th session, for each user, we build a graph where nodes correspond to target user and her friends. For target user $u$ with $|\mathcal{N}(u)|$ friends, the graph has $|\mathcal{N}(u)|+1$ nodes.   
we rerepresent nodes as $\boldsymbol{h}_{u}^{(0)}=\widetilde{\boldsymbol{u}}_{t-1}$ and $\{\boldsymbol{h}_{j}^{(0)}=\boldsymbol{s}_{j}\}$, where user node representation $\widetilde{\boldsymbol{u}}_{t-1}$ is gotten from last TIE module (see section \ref{www20_TIELAYER}). Then we employ relational-GAT over $\mathcal{N}(u)$. Specifically, we first project friends to the same space, and then calculate the attention score:
\begin{eqnarray*}
\label{CIKM2020_Industry_ratten}
\alpha_{uk}=\frac{\exp(f_{r}(\boldsymbol{h}_{u}^{(0)}, \boldsymbol{P}_{r}\boldsymbol{h}_{k}^{(0)}))}{\sum_{u_{j}\in\mathcal{N}(u)\cup \{u\}}\exp(f_{r}(\boldsymbol{h}_{u}^{(0)},\boldsymbol{P}_{r}\boldsymbol{h}_{j}^{(0)}))},\;\;\forall w\in\mathcal{N}(u),
\end{eqnarray*}
where $f_{r}(\cdot, \cdot)$ is the deep neutral network performing relational attention. Then, we aggregate information from $\mathcal{N}(u)$:
\begin{eqnarray*}
\boldsymbol{h}_{u}=\sigma\left(\sum_{u_{j}\in\mathcal{N}(u)\cup\{u\}}\alpha_{uj}\boldsymbol{h}_{j}^{(0)}\right),
\end{eqnarray*}
where $\sigma$ denotes the activation function. We denote the final representation of user as $\boldsymbol{u}_{t}=\boldsymbol{h}_{u}$.


\subsection{Temporal Information Encoding Module}
\label{www20_TIELAYER}

Temporal information is an important factor and there are traditional RS that consider temporal information \cite{DBLP:conf/wsdm/Song0WCZT19}. However, exploiting temporal personal preference and dynamic influences from users' friends together is still challenging. For example, traditional RNN usually performs worse as the length of the sequences increases. Besides in the real world, the influences produced in the earlier session may decay as time going by. Inspired by  Gated Recurrent Unit (GRU) \cite{DBLP:journals/corr/HidasiKBT15}
we propose the temporal information encoding modules. As shown in Figure \ref{www2020_framework}, we first select users’ historical behavior sessions, and utilize temporal information encoding modules among the outputs from each session to grasp the dynamic interests and influences. The output of relational-GAT module in session $t$ is denoted as $\boldsymbol{u}_{t}$. To combine relational-GAT encoded features at each time, as well as the target user's dynamic personal interests, we design a GRU-like module, called TIE module.
The hidden layer of TIE is a linear interpolation between the last TIE's hidden layer $\widetilde{\boldsymbol{u}}_{t-1}$ and the candidate hidden layer $\boldsymbol{\widetilde{h}}_{t}$. The encoding procedures are defined as below:
\begin{eqnarray*}
\setlength{\abovecaptionskip}{0cm}
\setlength{\belowcaptionskip}{0cm}
\begin{aligned}
\label{www20_TIE1}
\boldsymbol{u}_{q}=&\; \boldsymbol{W}_{q}^{t}\widetilde{\boldsymbol{u}}_{t-1}+\boldsymbol{b}_{q}^{t}\\
\label{www20_TIE2}
\boldsymbol{u}_{e}=&\;\boldsymbol{W}_{e}^{t}\boldsymbol{u}_{t}+\boldsymbol{b}_{e}^{t}\\
\label{www20_TIE3}
\widetilde{\boldsymbol{h}}_{t}=&\;\tanh(\boldsymbol{W}_{h}^{t}\boldsymbol{u}_{t}+\boldsymbol{u}_{e}\circ\boldsymbol{U}_{h}^{t}\widetilde{\boldsymbol{u}}_{t-1}+\boldsymbol{b}_{h}^{t})\\
\label{www20_TIE4}
\widetilde{\boldsymbol{u}}_{t}=&\;(\boldsymbol{1}-\boldsymbol{u}_{q})\circ\widetilde{\boldsymbol{u}}_{t-1}+\boldsymbol{u}_{q}\circ\widetilde{\boldsymbol{h}}_{t},
\end{aligned}
\end{eqnarray*}
where $\boldsymbol{W}_{q}^{t}$, $\boldsymbol{W}_{e}^{t}$, $\boldsymbol{W}_{h}^{t}$, $\boldsymbol{U}_{h}^{t}\in\mathbb{R}^{d\times d}$,  $\boldsymbol{b}_{q}^{t}$, $\boldsymbol{b}_{e}^{t}$, $\boldsymbol{b}_{h}^{t}\in\mathbb{R}^{d}$. 
We let $\widetilde{\boldsymbol{u}}_{0} = \boldsymbol{u}$, the initial user latent feature $u$. So the vector $\widetilde{\boldsymbol{u}}_{t}$ can be seen as user's long-term preference combining her evolving interests and the dynamic influences from her friends. 

\subsection{Prediction}
The final output of relational-GAT module $\boldsymbol{u}_{T}$ is used as the target user representation, then the user representation $\boldsymbol{u}_{T}$ and the target item $\boldsymbol{v}$ are combined to predict the clicking probability $\hat{y}_{uv}=\sigma(f(\boldsymbol{u}_{T},\boldsymbol{v}))$, where $\sigma(\cdot, \cdot)$ is sigmoid function, and $f$ is a ranking function which can be either a dot-product function or a deep neural network. The loss function $L$ is the sigmoid cross entropy loss: 
\begin{eqnarray*}
\sum_{(u,v)\in R}-\left(y_{uv}\log\sigma(f(\boldsymbol{u}_{T},\boldsymbol{v}))+(1-y_{uv})\log(1-\sigma(f(\boldsymbol{u}_{T},\boldsymbol{v})))\right)
\end{eqnarray*}

\section{Experiment}

To comprehensively study our proposed model DREAM, we conduct experiments on real-world datasets to answer the following questions: Q1: Does DREAM outperform the state-of-the-art baselines? Q2: How is DREAM affected by each component?

\begin{table}[!h]
\setlength{\abovecaptionskip}{0cm}
\setlength{\belowcaptionskip}{0cm}
\centering
\small{
\caption{Descriptive statistics of two datasets}
\label{www20_dataset}
\label{table1}
\begin{tabular}{c|c|c}
\toprule[1pt]
	             &Epinions  		&	Movie	\\
\hline
$\#$ Users	&	15,489  	&	57,496			\\
$\#$ Items	&	255,253 		&	56,858		\\	
$\#$ Events	&	500,770 		&	3,007,442		\\	
$\#$ Social links	&	355,217 		&	1,758,302		\\	
Avg.session/user  &2.3778          &2.8786    \\
Avg. real friends/user	&	22.99 		&	15.60 		 	\\
\bottomrule[1.0pt]
\end{tabular}
}
\end{table}

\subsection{Experimental Setup}

{\bf Datasets.} {\it Epinions$\footnote{http://www.epinions.com}$} and {\it{Douban-Movie$\footnote{http://movie.douban.com}$}} (Short for Movie) are utilized to evaluate the performance of  DREAM. As for Epinions, ``trust" endorsements are converted to directed edges in social network and users' behaviors are segmented into {\it month-long} sessions. As for Douban-Movie, we crawled the interaction data using the identities of users in movie community along with associated timestamps and users’ social netowrks. We construct our datasets by using each review as an evidence that a user consumed an item and segment users’ behaviors into {\it week-long} sessions for their high activeness.We convert users’ explicit ratings on items into $1$ as the implicit feedback for all the datasets. We randomly split the user-item interactions of each dataset into training set ($80\%$) to learn the parameters, validation set ($10\%$) to tune hyper-parameters, and testing set ($10\%$) for the final performance comparison. The statistics of two datasets are presented in Table \ref{www20_dataset}.

{\bf Baselines.} In this subsection, we compare DREAM with four groups of recommendation baselines. (1) only considers user feedbacks: BPR\cite{DBLP:conf/uai/RendleFGS09}. (2) considers social network information:
GraphRec \cite{DBLP:conf/www/Fan0LHZTY19} and SBPR \cite{DBLP:conf/cikm/ZhaoMK14}. (3) considers the sequence of user actions: GRU \cite{DBLP:journals/corr/HidasiKBT15} and SASRec \cite{DBLP:conf/icdm/KangM18}. (4) combines social network and user actions’ temporal information into recommendation task: DGRec \cite{6547630}.

We also construct several variants of DREAM for ablation study on two datasets. There are two special changes in our DREAM. One is the addition of virtual friends’ information in social network, the other change is the utilization of more than one session information (the best parameter is 2). As for inner-session information, we design two variants of DREAM, DREAM-R and DREAM-V to evaluate the effect of virtual friends' information. DREAM-R removes virtual friends' information, while DREAM-V uses virtual friends' information only. As for inter-session information, DREAM-GAT utilizes GAT\cite{DBLP:conf/iclr/VelickovicCCRLB18} to combine two friend-level information rather than realtional-GAT. And,DREAM-TGRU leverages GRU to fuse two session information to evaluate the necessity of TIE module. DREAM-1 uses only one session’s information like DGRec, DREAM-3 utilizes 3 sessions' information.

{\bf Metrics.} R@K (Short for Recall@K,K=10), NDCG and MRR are utilized to measure the performance. For all the metrics, the larger the values, the better the performance. In order to reduce the computational since there are too many unrated items, we randomly sample 1000 unrated items as negative samples and combine them with positive items in the ranking process for each user. We repeat this procedure 10 times and report the average ranking results.





{\bf Parameter Setting.} For all the models that are based on the latent factor models, we initialize the latent vectors with small random values. We use Adam \cite{DBLP:journals/corr/KingmaB14} as the optimizing method for all models that relied on the gradient descent based methods with a learning rate of 0.0001 and batch size of 32. In DREAM model, GloVe algorithm used to find high-quality virtual friends gets the best performance compared to rating matrix. The real and virtual friends sampling numbers are uniformly set to 10 after lots of experiments. For relational-GAT module in DREAM, we set the corresponding parameters according to their original paper \cite{DBLP:conf/cikm/XuLHLX019}. The regularization parameter as 0.00001 and ReLU is utilized to implement the non-linear transformation function. In order to make the models converge faster, we apply the batch normalization. The best models are selected by early stopping when the validation accuracy does not increase for 5 consecutive epochs. There are several other parameters in the baselines, we tune all these parameters to ensure the best performance of the baselines for fair comparison. 


\subsection{Model Analysis}

{\bf Overall Performance: Q1.} As shown in Table \ref{table3}, the following observations can be made: SBPR and GraphRec utilize both user-item interactions and social relations: while BPR only uses user-item interactions. SBPR and GraphRec outperform BPR, which is consistent with previous work. This indicates that social information reflects users’ interests effectively.

\begin{table}[!h]
\setlength{\abovecaptionskip}{0cm}
\setlength{\belowcaptionskip}{0cm}
\centering
\footnotesize{
\caption{Overall performance. ‘Imprv.’ denotes percentage improvement of DREAM, with respect to the best baseline.}
\label{table3}
\begin{tabular}{c|c|c|c|c|c|c}
\toprule[1pt]
 \multirow{2}{*}{Model}  &   \multicolumn{3}{|c|}{Epinions}	&\multicolumn{3}{|c}{Movie}\\
\cline{2-7}
 &  R@10  & NDCG  & MRR    &  R@10  & NDCG  & MRR \\ 
\hline
	BPR	&	0.00585 	&	0.08396 	&	0.00228 	&	0.01574 	&	0.11265 	&	0.00651 	\\
\hline
	SBPR	&	0.00658 	&	0.08948 	&	0.00281 	&	0.01642 	&	0.11333 	&	0.00685 	\\
	                                       GraphRec	&	0.00880 	&	0.09635 	&	0.00409 		&0.01787 	&	0.11352 	&	0.00698 	\\
\hline
	GRU	&	0.00410 	&	0.09229 	&	0.00360 	&	0.01141 	&	0.11380 	&	0.00700 	\\
	                                            	SASRec	&	0.00410 	&	0.09239 	&	0.00287 		&0.01723 	&	0.11459 	&	0.00747 	\\
\hline
	DGRec	&	{\bf 0.01176} &	{\bf 0.09632} 	&	{\bf 0.00468} 	&	{\bf 0.01901} 	&	{\bf 0.11486} 		& {\bf 0.00750} 	\\
	DREAM	&	0.01639 	&	0.09787 	&	0.00628 	&	0.02285 	&	0.11669 	&	0.00870 	\\
\hline
{\it Imprv.}			&	39.37$\%$	&	1.58$\%$	&	34.19$\%$	&	20.20$\%$	&	1.59$\%$	&	16.00$\%$	\\
\bottomrule[1.0pt]
\end{tabular}
}
\end{table}

In most cases, GRU and SASRec obtain better performance than BPR, which are modeled with temporal information. These improvements reflect the power of temporal information on RS. However, the performance of methods of social relations sometimes exceed temporal information based methods, which suggests it is hard to determine which is better. DGRec combining social relations and temporal information achieves much better performance than classic, social and temporal baselines. DGRec utilizes dynamic interests and static interests social information to represent target user’s interests, meanwhile, it employs graph-based algorithm with attention mechanism to deal with the influence of social friends, which shows the effectiveness of these two changes.

Our model DREAM consistently outperforms all the baselines. The substantial improvement of DREAM over the baselines could be attributed to three reasons: 1) We expand target user’s social network using virtual friends, which comprehensively expresses target user’s dynamic and static interests. 2) We combine user’s historical representation and current representation as the input of TIE. Utilizing updated user representation considers is the important step to learn the evolution of target user’s interests. 3) Different from DGRec only using one session information, we design TIE module to employ multiple temporal sessions’ information which reflects the evolution of target user’s dynamic interests over time. This strongly indicates the advantage of using temporal information across sessions. In terms of time complexity, we find that DREAM using 2-session information (25 seconds) increases 1.9 seconds than DGREC (23.1 seconds)  by calculating running time spent on single GPU per epoch.

{\bf Ablation Study: RQ2.} Table \ref{table9} are the ablation study's results of DREAM in terms of R@10 and MRR. For inner-session aspect, we compare DREAM with DREAM-R and DREAM-V. DREAM-V captures users' dynamic interest information hidden in graph structure, while DREAM-R makes up corresponding social information. So, DREAM using both real and virtual friends' information gets better performance than DREAM-R and DREAM-V. In inter-session aspect, we also evaluate the necessity of relational-GAT dealing with differernt kinds of social information and TIE module. The performance of DREAM-GAT is slightly poorer than ours, indicating relational-GAT can sharply catch the difference between the two social information. The results show that TIE module improves the performance compared with DREAM-TGRU.  This fully verifies the temporal effects in capturing users' interests. More importantly, the more session information doesn’t mean better performance. DREAM utilizing 2 session’s information gets better performance than DREAM-3.  Moreover, the more session information used, the longger running time it takes to run.

\begin{table}[!h]
\setlength{\abovecaptionskip}{0cm}
\setlength{\belowcaptionskip}{0cm}
\centering
\small{
\caption{Ablation study comparing the performance of the
complete model DREAM with several variations.}
\label{table9}
\begin{tabular}{c|c|c|c|c|c}
\toprule[1pt]
 \multicolumn{2}{c|}{Model} & \multicolumn{2}{|c|}{Epinions}		& \multicolumn{2}{|c}{Movie}\\	
 \cline{3-6}
 \multicolumn{2}{c|}{Components} &R@10	&MRR	&R@10&	MRR\\
 \hline
Inner-&	DREAM-R	&	0.00820 	&	0.00325 	&	0.01868 	&	0.00759 	\\
Session &	DREAM-V	&	0.01230 	&	0.00347 	&	0.01873 	&	0.00765 	\\
\hline
&	DREAM-GAT	&	0.01510 	&	0.00527 	&	0.02109 	&	0.00816 	\\
\cline{2-6}
Inter-&	DREAM-TGRU	&	0.01530 	&	0.00551 	&	0.02186 	&	0.00837 	\\
\cline{2-6}
Session&	DREAM-S1	&	0.01297 	&	0.00389 	&	0.01931 	&	0.00749 	\\
&	DREAM-S3	&	0.01430 	&	0.00486 	&	0.02000 	&	0.00826 	\\
\hline
 \multicolumn{2}{c|}{DREAM}			&	0.01639 	&	0.00628 	&	0.02285 	&	0.00870 	\\
\bottomrule[1.0pt]
\end{tabular}
}
\end{table}

\section{Conclusion}

In this paper, we propose DREAM for social RS. DREAM tries to model both users’ dynamic interests and their friends’ temporal influences. Specifically, in each session, to solve the sparsity of social relations, we design a GloVe-based method to increase the number of friends, and utilize relational-GAT to integrate influences from friends. And then we build TIE modules to encode the outputs form historical sessions by recursively combining the features encoded by relational-GAT modules and that from last TIE module. By doing so, the user representations involve both the user's dynamic interests and the dynamic influence from her friends. In the extensive experiments on the public datasets, DREAM significantly outperforms the state-of-the-art solutions.

\bibliographystyle{ACM-Reference-Format}
\bibliography{DREAM}

\end{document}